\documentstyle[twoside,fleqn,espcrc2]{article}

\newcommand{\AmS}{{\protect\the\textfont2
  A\kern-.1667em\lower.5ex\hbox{M}\kern-.125emS}}

\title{Random geometries and real space renormalization group}

\author{D.A. Johnston\address{Mathematics Department, Heriot-Watt University,
Edinburgh, Scotland}
 , J.-P. Kownacki$^{\rm b}$
 and
A. Krzywicki\address{LPTHE, B\^{a}timent 211, Universit\' e Paris-Sud, 91405
Orsay, France}\thanks{Laboratoire associ\' e au C.N.R.S.}}
\begin{document}

\begin{abstract}
A method of "blocking" triangulations that rests on the self-similarity feature
of dynamically triangulated random manifolds is proposed and used to define the
renormalization group for random geometries. As an illustration, the idea is
applied to pure euclidean quantum gravity in 2d. Generalization to more
complicated systems and to higher dimensionalities of space-time appears
straightforward.
\end{abstract}

\maketitle

\par
We shall report on a work which is still in progress.
The aim of this contribution is to share with you
the underlying idea.
\par
Although the recent advance in the statistical
mechanics of discrete random
manifolds has been
quite impressive, it appears that our ability to
simulate complicated models on a computer exceeds
our ability to understand the real significance of
the underlying physics (except for a class of 2d models). Contrary to discrete
field
theories where the lattice is only an inert scaffolding,
in the case of discrete gravity it is not always evident
how to fix the physical scales and how to define the continuum limit. The
answers to such questions are
usually provided by the renormalization group. One
apparently needs to supplement the existing computer artillery with the
techniques of the real space renormalization group (RG). In the present context
this
requires inventing some analogue of Kadanoff's blocking \cite{kadan} applied to
the geometry itself. One
suggested approach has been to let coarser versions
of an initial triangulation ``follow''
the original lattice with
some appropriate rules as the connectivity
is changed during
a simulation \cite{renken}. In this work we take a
different tack, expanding on the
ideas put forward in ref. \cite{krz}. The dynamical triangulation recipe is
adopted.
\par
Conceptually, there are two elements in the familiar
Kadanoff construction:
\par
(a) define larger geometrical cells with respect to the smaller ones.
\par
(b) define "block" spins on larger cells in terms of the spins living on the
smaller cells.
\par
On a regular lattice (a) is trivial: the lattice
obtained by removing, say, every 2nd point in each
direction is identical modulo rescaling to the original lattice. A change of
scale is automatically a
self-similarity transformation. On a random lattice
(a) is a problem in itself and requires some thought. Therefore, we focus on it
here, leaving aside (b) which
does not seem to present any serious conceptual difficulty.
\par
We observe that the essential ingredient of the
Kadanoff construction is a similarity transformation
of the geometry, rather than mere decimation of lattice points (remember that
values of critical
couplings are lattice dependent).
Another important remark is that a randomly
triangulated surface is a fractal. This has been
emphasized by many people, in particular by Jain and
Mathur \cite{jm} in their paper on baby universes (BUs),
which is particularly relevant to this work. Our
idea consists in using this self-similarity to
{\em define} the step (a). We introduce
one additional piece
of terminology with respect to \cite{jm}, namely:
BUs with no further BUs growing on them are called
{\it last generation} BUs.
The surface is a hierarchical
object so cutting last generation BUs one gets the same
ensemble of surfaces modulo rescaling.
\par
Consider 2d gravity for simplicity. There is an
 immediate problem: cutting BUs with neck of
length $l$ one creates a polygon which is not
necessarily a triangle (it is a triangle when
$l=3$, only). This is similar to the problem
one encounters in Br\'ezin-Zinn-Justin
renormalization group program \cite{bzj},
where starting from
$\phi^3$ one generates all interactions. Thus
it would be natural, here also, to consider
surfaces made out of arbitrary polygons.
Generically, such an ensemble of surfaces is
expected to belong to the same universality
class as the ensemble of triangulated surfaces,
i.e. pure 2d gravity. However, although the
extension of the model to general cellular
decompositions of surfaces does not present
any special difficulty, the ``BU surgery''
becomes rapidly more difficult to implement when
the neck length $l$ is larger than 3.
\par
Fortunately, it seems completely sufficient to
limit one's attention to BUs with minimum neck
length $l = 3$, to be called minBUs following
the terminology introduced in ref. \cite{jm}.
This is the choice made in this exploratory paper.
\par
In practice, one must distinguish BUs from a
smooth deformation of the surface setting a
lower limit on the number of points $B$
in a BU: $B > B_0$~. It turns out that relevant
results are insensitive to the choice of $B_0$.
\par
Define $\Delta = \mu_c - \mu$, where $\mu$ is the cosmological constant and
$\mu_c$ is its critical
value. Consider an appropriate derivative of the
partition function $Z(\Delta)$, to be denoted
$\chi(\Delta)$ such that at small $\Delta$

\begin{equation}
\chi(\Delta) \sim \int dA A^{\Gamma-1} e^{A \Delta}, \; \Gamma > 0
\label{chi1}
\end{equation}

\noindent
The condition $\Gamma > 0$ insures that
the integral diverges when $\Delta \to 0$.
In generic topology $\chi(\Delta)$ is the
so-called string susceptibility and $\Gamma$
is the corresponding critical exponent,
conventionally denoted by $\gamma$. In spherical
topology, we take an extra derivative of
$Z(\Delta)$, so that $\Gamma = \gamma +1$.
Neglecting non-singular contribution to the right-hand side one has

\begin{equation}
\chi(\Delta) \sim \Delta^{-\Gamma} , \; \Gamma > 0
\label{chi2}
\end{equation}

\noindent
For pure 2d gravity and spherical topology
one has exactly $\Gamma = 1/2$.
\par
We shall now present the results of a computer experiment implementing and
illustrating our ideas. We work
in spherical topology.
\par
Cutting minBUs produces a mapping of the
grand canonical ensemble of surfaces into
itself. We first determine the area distribution
of the ensemble of surfaces obtained by cutting
last generation minBUs growing out of a (randomly
chosen) surface with {\em fixed} number
 of points $A_0$, in
a sense the ``image'' of a ``point source''. We
find that this image is a nearly Gaussian curve

\begin{equation}
I(A \mid A_0) = {{1} \over \sqrt{2\pi\sigma^2}}
e^{-(A - \lambda A_0)^2/2\sigma^2}[ 1 + ...]
\label{gauss}
\end{equation}

The parameter $\lambda$ becomes rapidly independent of
$A_0$ and $\sigma^2$ increases
linearly  with $A_0$. Higher order cumulants,
once scaled by an appropriate power of $\sigma$,
are small and seem
to decrease with $A_0$.
\par
Our experiment has been run with dynamically
triangulated surfaces,
the number of triangles
ranging from 1000 to 20000. For each value of
$A_0$ we have carried out a series of several
dozens mini-experiments, each mini-experiment
corresponding to typically 5000 sweeps of the
lattice. About 1000 heating sweeps have been
performed between two mini-experiments. The
measure of $A$ has been done once every 10 sweeps.
The errors have been estimated using the standard
binning method.
\par
The exact numerical values of the parameters
$\lambda$ and $\sigma$ depend on the choice
of the lower cut-off $B_0$, defining an
``acceptable'' minBU. In most of our calculations
we have set $B_0 = 10$. An excellent
fit to the data is then obtained with
$\lambda = 0.8189(2) - 7.4(4)/A_0$,
$\sigma^2 = 2.42(2) A_0 + 185(45)$. Decreasing
(increasing) $B_0$ one gets, of course,
smaller (larger) $\lambda$. For example,
with the choice $B_0 = 15$ one finds
$\lambda \approx 0.859$. The correction
terms in (\ref{gauss}) can be
organized in a Gram-Charlier series
\cite{kend}. A very good description of
the image is obtained including the first
 two correction terms in this series.
The  values of the third and
fourth order cumulants
for various $A_0$ are given
in Table 1. The (scaled) cumulants are
small and the deviation from the Gaussian
is not large, especially at large $A_0$.

\begin{table}[hbt]
\setlength{\tabcolsep}{1.5pc}
\newlength{\digitwidth} \settowidth{\digitwidth}{\rm 0}
\catcode`?=\active \def?{\kern\digitwidth}
\caption{Scaled third ($\kappa_3$) and fourth ($\kappa_4$) order cumulants
against $A_0$}
\label{tab:cumulants}
\begin{tabular}{ccc}
\hline
$A_0$ & $\kappa_3/\sigma^3$ & $\kappa_4/\sigma^4$   \\
\hline
1000 & -0.50(3) & 0.70(14) \\
2000 & -0.33(2) & 0.28(3)  \\
3000 & -0.24(7) & 0.17(11) \\
4000 & -0.22(2) & 0.11(5)  \\
7000 & -0.11(7) & 0.12(11) \\
8000 & -0.21(5) & 0.03(13) \\
9000 & -0.21(5) & -0.01(10) \\
10000 & -0.11(4) & 0.02(6) \\
\hline
\end{tabular}
\end{table}

The image of the full ``source''
$S(A_0,\Delta) \sim A_0^{-1/2} \exp{(A_0 \Delta)}$ is
given by the convolution

\begin{equation}
{{S'(A, \Delta)} } = \int dA_0 \;
I(A \mid A_0) \;  S(A_0, \Delta)
\label{trans}
\end{equation}

\noindent
and can be calculated analytically for large $A$,
using the saddle-point method. Keeping only the
first term in (\ref{gauss}) and remembering that
$\sigma^2 \sim c A_0$ one finds for $\Delta \ll 1$

\begin{equation}
S'(A, \Delta) \sim {{1} \over \sqrt{\lambda A}}
e^{A \Delta/\lambda}
\label{res}
\end{equation}

\noindent
It is remarkable that the right-hand side does
not depend, for small $\Delta$, on the value of
the constant $c$. Since the image $S'(A, \Delta)$
and the source $S(A_0,\Delta)$
are similar, integrating eq. (\ref{res})
with respect to $A$ gives
the scaling law

\begin{equation}
\chi(\Delta) =  \lambda^{-\Gamma}
 \chi(\lambda^{-1}\Delta ) , \; \Gamma = 1/2
\label{scaling}
\end{equation}

\par

Since scaled  cumulants do not seem to grow
with $A_0$,
the term proportional
to the ${\rm n}^{\rm th}$ order Hermite
polynomial in the Gram-Charlier series
contributes a correction of maximal
order $\Delta^{(n-1)/2}$ to the
image $\chi(\Delta)$. Therefore, the
corrections do not contribute to the singular
part of the image (at least treated term by term).
\par
Although the operation of cutting
last generation minBUs
is not just a  rescaling of the area, we
find that the image of a point source
is sufficiently sharp for the blocking
operation to implement a homogeneous
transformation of $\chi(\Delta)$, as long as
one is close enough to the critical
point. It is clear that the repeated
application of this operation defines
a renormalization group, in the usual
way, so that $\lambda$ in (\ref{scaling})
can be replaced by an adjustable scaling
factor $\Lambda = \lambda^n$.
\par
We expect (\ref{scaling}) to hold also
for other topologies, when $\Gamma$
differs from 1/2. Obviously, the effective
$\Delta \to 0$  for
$\Lambda \to \infty$. A fully fledged
fractal is obtained starting
with a smooth surface and repeating the
operation of {\em adding} minBUs infinitely
many times. This corresponds to the limit
$\Delta \to 0$~: being  fractal and critical
is synonymous.
\par
Generalization to $d > 2$ seems conceptually
straightforward. In the simplest
model one has two independent
couplings, the cosmological and
the Newton constants, respectively.
Thus, instead of the single
$\Delta$, there are two such parameters.
One can therefore fix independently
two dimensionful constants (as it should
be for gravity).
\par
In conclusion, we propose how to use
the fractal, hierarchical structure of
the euclidean space-time foam
to define a real space
renormalization group for random
geometries. It has been important to
check that the idea is meaningful for
surfaces of accessible size. In our current work we
examine various applications of the
idea presented above, focusing on a
search for most suitable observables
to define the $\beta$ function.

\end{document}